\documentclass[11pt]{article}
\usepackage{times}
\usepackage{amsmath,amsopn,amscd,amssymb,xypic,rotating,newapa}
\xyoption{all}

\newcommand{\separate}    {\vspace{0.3cm}\begin{center}*~~~~~~~~~~*~~~~~~~~~~*\end{center}\vspace{0.3cm}}

\newcommand{\cb}          {\begin{tabbing}MMMMM\=MM\=MM\=MM\=MM\=MM\=MM\=MM\=MM\=MM\= \kill}
\newcommand{\ce}          {\end{tabbing}}

\newcommand{\double}       {\baselineskip 20pt}

\newcommand{\dqt}[1]        {``{#1}''}

\newcommand{\VInsert}[2]   {{\immediate\pdfximage height #2 {#1}\pdfrefximage\pdflastximage}}

\renewcommand{\topfraction}{.99}
\renewcommand{\textfraction}{.01}

\begin{document}

\double

\title{SEMIOTIC INTERNATIONALIZATION AND LOCALIZATION OF COMPUTER PROGRAMS}
\author{Simone Santini}
\date{Escuela Polit\'ecnica Superior\\Universidad Aut\'onoma de Madrid\\simone.santini@uam.es}

\maketitle

\begin{abstract}
Localization, the process--part of translation studies--of adapting a
program to a new linguistic community, is often intended in the
relatively narrow sense of translating the messages and labels of the
program into the target language. Correspondingly,
internationalization, the discipline--which is part of software
engineering--of putting in place all the measures that will make
localization easier, is also limited in scope.

In this paper we analyze the various systems through which a program
communicates with a person (icons, buttons, actions, interface layout,
etc.) and find that most of them, far from being iconic, are in
reality symbolic semiotic systems related to the culture in which or
for which the program was developed (typically American programmers of
western office workers). Based on these findings, we argue that during
the localization process, the translator should have the option to
translate them all, that is, to adapt the whole interface and its
founding metaphors to the cultural environment in which the program is
deployed.

This conclusion will result in a greater r\^ole for
internationalization in the software development process, and we
outline a few architectural principles that should be considered when
creating a program for a multi-cultural market.
\end{abstract}

\section{INTRODUCTION}
\emph{Localization} is an area of translation whose purpose is to
adapt computer programs (or, more precisely, the programs' interfaces)
to the various linguistic and cultural environments in which they will
work. Related to localization is \emph{internationalization}: a family
of design practices whose purpose is to make localization feasible and
as painless as possible \cite{uren:93,cooper:01,cyr:04}.  Localization
and internationalization are often mentioned together and they are,
indeed, closely related, but they are not the same thing. Not by a
long shot. They are not even part of the same discipline: localization
is related to anthropology and translation studies, while
internationalization is a quintessential software engineering
problem. While this division is such that software engineers will work
essentially in internationalization, their work is destined to offer a
support to localization, so it is necessary that they be aware of the
problems and charachteristics for which they are offering support.

Localization is often seen in a fairly narrow sense, as the
translation of the various captions and labels that appear in a
program into a \emph{target language}, and internationalization is
correspondingly limited to the engineering practices that make this
translation easier: reserving spaces in labels and buttons, using full
phrases as the display units, making sure that the program works well
with a variety of alphabets and writing directions, and so on.  As an
example of this attitude, out of the 49 principles for
internationalization given in \cite{hall:02}, only two are not
directly related to language:
\begin{description}
\item[i)] icons, cursors, and bitmaps are generic, are culturally
acceptable, and do not contain text;
\item[ii)] if ethnocentric graphics, colors, or fonts, are used, they
can be replaced dynamically using locale-sensitive switch statements.
\end{description}
This emphasis on the linguistic aspects of localization is,
\emph{prima facie}, sensible: language is pervasive, cultural,
symbolic, and its signifiers are opaque \cite{hodge:88}.  A program
message written in Chinese will provide absolutely no clue to a French
reader (unless the reader happens to read Chinese), and vice-versa.
Also, even in an age of graphical interfaces and touch screens,
language remains the main instrument of communication between a
programmer and a user, and few programs can be profitably used by
somebody who doesn't understand the language in which its messages are
expressed.

Still, language is not everything. A program and its users communicate
through a collection of semiotic systems of which language is but one.
The most preminent of such systems, in modern computer programs, is
that of \emph{icons}. Superficially, one might regard icons as the
prototypical example of transparent signifiers (iconic, in fact) and,
consequently, one might consider that they are in no need of
translation. Transparent signifiers work through non-arbitrary
associations with their signifieds \cite{merleau:64}; icons, in
particular, signify by means of a similarity between signifier and
signified, and this similarity is supposed to be
cross-cultural. Universal, in fact.  I will argue that this is not the
case: computer icons, their name and of their aspect of little images
notwithstanding, are not iconic; rather, they form a sophisticated
symbolic code that, as all symbolic codes, works by
mutual consent of the cultural community for which it
functions. Icons, in other words, are a language no less symbolic than
English or Malawi, and no less in need of translation than
these. Consequently, I shall argue that the programme of localization
should be much more ambitious than it has been heretofore proposed,
and that it should involve the cultural adaptation of all the
non-transparent semiotic systems that compose a computer interface,
from language to icons, and of the actions that the user performs on
them.

This is a new task for localization and, especially, for
internationalization, because this kind of \dqt{total semiotic
  translation}, so to speak, requires the possibility to reconfigure a
program at a much deeper level than that required by the standard
acceptation of localization, and exerts an influence on the design
process that goes well beyond a simple list of \emph{desiderata} such
as Hall's.  Internationalization, in this view, becomes a much more
relevant discipline within software engineering, one that must
profoundly influence the very structure of a program, and that should
become an integral part of a design from the very beginning.

Modern interfaces are often designed based on a metaphor, that is, on
the (more or less complete) analogy between the symbols and operations
of the interface semiotic system and those of a second system with
which the user is assumed to be familiar.  We call the former the
\emph{manifest system}, and the latter the \emph{anchor system} of the
interface. In computers, the manifest system is subject to a certain
amount of variation: not only does the manifest semiotic system of,
say, a computer with a MacOS operating system look somewhat different
from that of a computer with a Linux operating system, but also some
portion of the manifest system of, say, a word processor will be
different from those of a mathematics program (for instance, the same
symbol may mean different things in the two contexts, or be placed in
another differential relations with the other symbols of the
interface).

The \emph{anchor system} is often the same, and it has not changed
since the technicians at Xerox's PARC research center created the
interface for the ALTO computer in the 1970's \cite{lavendel:80}: an
office desk with documents organized in folders. From the localization
standpoint, it is important to determine whether this metaphorical
substratum is sufficient to make the signifiers of the manifest system
transparent: if it is not, then the manifest system is opaque, and
must be localized when the program is adapted to a new culture.  For
the system to be transparent, two conditions must be met:
\begin{description}
\item[i)] the relation between the manifest system and the anchor
  system must be iconic, that is, the actions performed on the
  interface must correspond iconically to the actions typical of the
  anchor system, and
\item[ii)] the anchor system must be known to the cultural group for
  which the program is adapted, so that the actions pointed to by the
  elements of the interface can be recognized.
\end{description}
I claim that none of these conditions is verified in a typical
computer interface. On the one hand, the icons stand for certain
actions only conventionally and are therefore symbols.  On the other
hand,the \dqt{point and click} interface was born in an environment of
technicians who worked in offices, and while the desk-and-documents
anchor system was very familiar to the office-working engineers of
Silicon Valley, the same is not necessarily true for other users of
computer systems.  The applications that these technicians were
desigining were intended as an aid to office workers, and considering
computer files as documents made good sense in this context. Today,
however, the data that are stored in a computer are not of the nature
commonly found in offices (just think of videos and songs), and the
people who are using the computers are not necessarily familiar or
comfortable with office work. The metaphor has lost its force and its
\emph{raison d'\^etre}. Because of this, the graphical elements and
the actions of an interface are just as symbolic as the text that the
interface contains, and equally in need of translation in order to be
adapted to the different cultural communities to which the program is
directed.  This requires a fundamental change in the way one designs
interfaces. It requires that the interface to an application be
designed at a much more abstract level, using eitehr a
non-metaphorical language of algebraic operations, or a metaphor with
the broadest diffusion: in this paper I propose the use of
\emph{conversation} as a metaphor for the anchor system. The manifest
system should be as independent as possible of the rest of the
program, easily repleaceable and configurable during the use of the
program.  It should be part of the mission of internationalization to
facilitate this change.  Ideally, the manifest system should be
determined by a series of configuration parameters in the application
so that a word processor designed for office workers (with the
interface based on a certain model and with certain assumptions) can
be changed quickly in an application for college students (with a
different manifest interface based on a different set of
assumptions). The same application, running on the same computer,
should be able to present completely different manifest systems to
different users.  The interface, in other words, should be a
collection of configurable, independent modules, connected to the
application in a standard way, but functionally independent of it.

\section{METAPHORS WE PROGRAM BY}
The anchor system that we use, and the metaphor that it entails are
not simple or neutral issues; far from being just a matter of
convenience, they determine the way in which we think.  Our everyday
language is filled with metaphors (\emph{vide} the expression
\dqt{filled with metaphors} that I just used, which is based on the
metaphor \emph{language is a container}), and the metaphors that we
use determine to a considerable extent the limits of our possibilities
of thinking about the world.  Consider the metaphor \emph{debate is
war}, which forms the basis of expressions such as \dqt{winning an
argument}, \dqt{I rebuked his objections}, \dqt{when he said that, I
changed my strategy}, and so on. This metaphor comes with a set of
assumptions that determines how we conduct a debate, what counts as a
debate, how to behave in a debate, and so on \cite{lakoff:80}.

The metaphor allows us certain ways of acting in a debate, and
prevents others. Outside of the metaphor, for example, the idea that a
debate must have a winner is not at all obvious: one might think of a
debate as a way of clarifying a difficult topic by expressing
different opinions about it. In this second view, the ideas that there
should be a winner of a debate, that one participant should feel
compelled to defend one point of view, or to rebuke the arguments of
the other would be quite ludicrous.

Metaphors are cultural, and different cultures apply different
metaphors to the same topics. Consider, for example, the way people
see the changes that occur in their lives and in the society in which
they live with the passage of time. In our culture, dominated since
the enlightment by the idea of \emph{progress}, these changes are
perceived as a movement, specifically a movement towards a desirable
ideal placed in the future. Change is motion, we are walking from the
past to the future, and our gaze points in the direction in which we
are walking, so the future is \emph{in front} of us, and the past is
\emph{behind} us. When we are at the end of a difficult period, we
leave our problems \emph{behind us}, we let \emph{bygones be
  bygones}, and we \emph{move on} (or, at least, some of us do). In
this society, an expression such as \dqt{as time goes by} implies a
surrender to the flow of time, a renounciation to fight, to determine
one's life, a surrender to the event. It is altogether interpreted as
a negative and sad feeling. Something quite fitting for the central
musical theme of \emph{Casablanca}.

More contemplative societies might apply the opposite metaphor. We can
see the past, but the future is obscure, so the past is in front of
us, while the future is behind us, hidden from view. In this case,
change might not be a movement at all, but a static contemplation of
things as they go by. Things are initially behind us, then they come
into view and we see them, first clearly, then, as they start moving
farther away into the past, more dimly. The consequence of this
metaphor might be important: a society that perceive time as a motion
of the subject is more likely to assume that we can control where we
are going, and try to devise means to control and predict the future
(which amount to the same thing, since we often predict for the sake
of controlling). A society that looks at the past and sees time as
coming by is likely to be more fatalist, probably more religious or
spiritual, and will probably develop tools to deal with the unexpected
and the unpleasant (the concept of God's will, for example). The song
\emph{As time goes by} would not seem to them sad at all: it would
describe a normal, positive outlook on life.

Just like language metaphors are instrumental in establishing our
relation to the subject of our thoughts and in determining, in a way,
the orthodoxy of thought---the shape that our ideas are allowed to
take---so interface metaphors are instrumental in determining our
relation with an information system, not only in the sense of
constraining the actions that we are allowed to undertake (any
interface will do that) but, more importantly, in directing the way we
think about an activity and determining what we can conceive of doing.
A computer interface creates a doubly indirect relation between the
operations of a program and the elements that represent it on the
screen, a relation mediated, on the one hand, by the anchor metaphor
and, on the other hand, by the screen element that represents the
metaphorical action, as in the following schema:
\begin{center}
\setlength{\unitlength}{0.1em}
\begin{picture}(300,60)(0,0)
\put(0,10){
  \put(10,25){\line(0,1){5}}
  \put(60,25){\line(0,1){5}}
  \put(10,30){\line(1,0){50}}
  \put(35,40){\makebox(0,0)[b]{manifest}}
  \put(35,32){\makebox(0,0)[b]{system}}
}
\put(110,10){
  \put(10,25){\line(0,1){5}}
  \put(70,25){\line(0,1){5}}
  \put(10,30){\line(1,0){60}}
  \put(40,40){\makebox(0,0)[b]{anchor}}
  \put(40,32){\makebox(0,0)[b]{system}}
}
\put(36,22){\makebox(0,0){screen}}
\put(36,13){\makebox(0,0){element}}
\put(150,23){\makebox(0,0){metaphorical}}
\put(150,13){\makebox(0,0){action}}
\put(263,22){\makebox(0,0){program}}
\put(263,13){\makebox(0,0){action}}
\put(90,20){\vector(-1,0){30}}
\put(90,20){\vector(1,0){30}}
\put(210,20){\vector(-1,0){30}}
\put(210,20){\vector(1,0){30}}
\put(90,26){\makebox(0,0){represent}}
\put(90,12){\makebox(0,0){(iconic?)}}
\put(210,26){\makebox(0,0){represent}}
\end{picture}
\end{center}
The iconicity hypothesis of computer interfaces---that underlies the
assumption that non-verbal interface elements should not be
translated---rests on three hypotheses:
\begin{description}
\item[i)] the screen elements (elements of the manifest system) are
  transaprent iconic signifiers of actions and elements of the anchor
  system; for example, the picture of a folder \dqt{stands} iconically
  for a folder in the office desktop system (the anchor system of the
  interface), the paper basket stands for an actual paper basket% ,
  etc.%
  \footnote{The paper basket is a good example of the relation between
    the manifest system and local culture, and a clear indication of
    the relation between icons and the evolution of western
    culture. In the early computer the \dqt{delete} icon used to
    represent a metallic trash can; in the 1980's, years in which
    science was very fashionable, the NeXT compter had a black hole in
    which the documents would disappear; in the 1990s, more
    ecologically conscious, the device was transormed into a recycling
    bin, with a prominently visible recycling sign.}%
\item[ii)] the anchor system is isomorphic to the algebra of the
  program in that there is a one-to-one correspondence between actions
  in the anchor system (eg. throwing a piece of paper into the paoer
  basket) and actions of the program (e.g. erasing a document from the
  disk);
\item[iii)] the anchor system is one with which the intended user of
  the program is familiar, so that it can accomplish its function of
  mediating the interactions of the user with an unfamiliar program,
  and with the unfamiliar activities performed by that program.
\end{description}
In the following sections I shall try to problematize these
hypotheses, byt showing that they are by no means more obvious than
their opposites.  If these hypotheses fail, so does the assumptions on
the iconocity/metaphoricality of the connection between interface
elements and program actions, revealing that such a connection is
symbolic (or, at most, if one considers that the syntactic structure
of the interface is too simple to support symbols, indexical
\cite{deacon:98}). Once we have recognized that all interface elements
(and not only labels and captions) are symbolic, it follows that
localization must take care of the semiotic-cultural adaptation of all
interface elements, be they linguistic or non-linguistic. The whole
interface needs translation, not just its labels.

This form of \dqt{semiotic translation} entails not only an extension
of the usual contept of localization, but also a re-definition and
expansion in scope of the internationalization substratum for it. The
last sections of this paper will briefly describe some possible
directions of these activities and a few of the foreseeable technical
issues that their expansion will generate.

\section{ARE ICONS ICONIC?}
The working of the manifest system depends on a number of properties,
some of which are structural (viz. there should be an isomorphism
between the operations of the manifest system and those of the anchor
system), and others are immanent properties of the elements. I shall
look at the structural properties in the next section while, in this
one, I shall consider the basic question related to the intrinsic
properties of the element, one of great importance for localization:
whether icons are transparent signifiers (and therefore in no need of
translation), or whether they are opaque (and therefore, they do need
translation). In Peircean terms, icons are transparent signifiers,
since they are \emph{motivated} by an iconic ground
\cite{sonnesson:99}, that is, by an evident similarity with their
signifier, whereas symbols are opaque, since they are conventional%
\footnote{Things are not as sharply cut as this classification might
  lead one to think, though: even the symbolic system \emph{par
    excellence}, language, can be considered completely opaque only in
  a synchronic way. Diachronically, there are evolutionary effects
  that modify every language to make it more suitable for very common
  practical needs and, therefore, towards more transparency. In most
  languages, for instance, the words for \emph{yes} or \emph{no} are
  very short, a transparent signifier of their relevance.}%
 . The point can therefore be driven by the somewhat ironic question:
\dqt{are icons iconic?}. The answer is that, at least on a computer
screen, they are not. Computer icons are highly stylized symbolic
elements of a somewhat simplified language and, in most cases, they
bear little or no resemblance to the original metaphorical operation
that they were supposed to illustrate. Much like the iconic signifier
of the head of an ox eventually became the highly stylized letter
\dqt{A,} which bears virtually no trace of its iconic origin, so the
icons on a computer are part of a pictorial code that forms, to all
practical purposes, a system of opaque signifiers.

\begin{figure}
  {\centerline {\immediate\pdfximage height 1cm {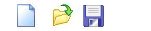}%
      \pdfrefximage\pdflastximage%
    }
  }
  \caption{Three icons commonly used in many computer programs that
    operate with documents. The first icon is associated to the
    function that \emph{creates} a new document, the second to the
    function that \emph{opens} an existing document, and the third to
    that which \emph{saves} on disk the document one is currently
    working on.}
  \label{icons}
\end{figure}

Let us consider, by way of example, the three icons that most users of
modern operating systems find on the upper-left corner of many
applications that they use, and that are reproduced in
figure~\ref{icons}.
The first icon shows a white rectangle with a corner cut off and a
triangle of the same size as the cut drawn just below it. The icon is
meant to give the idea of a blank sheet of paper, and the triangle
represents a folded corner of the sheet, pointing (iconically) to the
fact that the rectangle is made of a flexible material, such as paper,
and is not, say, a piece of wood or a brick. The iconic component here is
relatively strong, although it is doubtful that many people unfamiliar
with the world of computers or offices would immediately recognize
this picture as a piece of paper.  \emph{Once it has been pointed out
to them} that the icon represent a piece of paper with a folded
corner, people will have no problem recognizing its iconic
resemblance, but before this is pointed out to them, the icon is not
that clear. In this sense, the symbol is no more iconic than, say, the
Chinese ideogram \emph{k\`un} (\VInsert{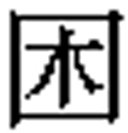}{1em}: weary, tired), which
represents a tree that grows in an enclosed space.

Once the ideogram has been explained, one has little trouble seeing
its iconic origin (as well as the poignantly poetic nature of the
semantic association) but, before the explanation comes, we are in the
presence of a fairly opaque signifier. In the case of the piece of
paper on the computer screen, the iconic nature might be a bit more
direct, but the signifier still qualifies, at best, as
semi-transparent.  The connection between the icon and the action is
also conventional. The icon is the representation of the action
\dqt{create a new document}, an action that, of course, exists only in
a computer: in the true embodiment of the anchor system (the office)
people do not create new documents \emph{ex nihilo}, rather they grab
a (existing) piece of paper and start writing.  So, while the relation
between the signifier and its physical referent (a piece of paper) is
semi-transparent, the relation between the icon and the action is
mediated by the highly stylized and highly symbolic nature of the
action, and is opaque.

The second icon is the visual representation of the command that
allows one to access an existing document for editing purposes. The
icon represents a common manila folder partially opened, sometimes
with a bent arrow indicating the movement of the opening folder, and
the associated program action is that of \dqt{opening} a
document. This icon has a curious history, moving at the same time in
the direction of a greater iconicity and of a more symbolic relation
to its referent. The movement towards a greater iconicity is due
largely to the greater graphical possibilities offered by faster CPUs
and graphic video peripherals. In its early versions, such as in the
Xerox's Alto, the icon was highly stylized, being essentially a
rectangle with a protuberance on top representing the protruding
label-tab that one finds in offices folders throughout the USA. As the
processing speed of personal computers increased, the icon has
acquierd color and depth and it is today easily recognizable as a
manila folder, at least to an American office worker. It must be
noted, however, that the MacOS operating system steps somewhat away
from the icon by making (in the default color scheme) the icon
blue. This is done in order to be consistent with the general
\dqt{cold} color scheme of the interface, and the operation is
possible precisely because the picture is not an icon but a symbol:
its meaning doesn't come from its similarity with a manila folder, but
from a differential relation with the other elements of the
interface. Changing the color, and reducing iconicity, doesn't change
the relation and, therefore, doesn't change the symbolic meaning of
the folder icon. This flexibility would not have been possible had the
folder been a true icon: reducing the similarity between the picture
and the real manila folder would, in that case, have reduced the
signifying power of the picture. To this we should add that the iconic
component of the symbol is cultural and, therefore, not
universal. Only certain groups of people will be familiar with this
kind of folders, namely certain office workers. The label tab, the
most distinguishing characteristic of the folder picture, is present
in actual folders only in some countries (notably in the US), while it
is virtually unknown in others (in Europe, for instance). This
cultural connotation, of course, reinforce the need to translate this
element.

In the context of a word processor the relation between the icon and
its signified action is symbolic not for want of realism, but because
the referent of the icon (the manila folder) is connected symbolically
to the action. The folder is partially open and has an arrow that
indicates the action of \emph{opening} a document (the use of an arrow
to indicate movement is, in itself, an opaque signifier related to the
symbolic language of comics \cite{varnum:01}), but the action that it
represents in the anchor system (viz. opening a folder) is not the
action that it is actually performing: on an actual desk one does not
\emph{open} a sheet of paper in order to change its contents, rather,
one grabs it, moves it, uncovers it, or does whatever action is
necessary to bring it in a place within the reach of hand and pen.
The connection between the word \emph{open} and the action of editing
a document is limited to computer practice and it is, at least from an
user point of view, arbitrary
\footnote{The name of the action comes from the computing term
  \emph{opening a file}, which is the action that a program does
  before it can access the contents of a file. The action is reflected
  at various levels in a computer. For instance, a program written
  using the C programming language must call the function \emph{fopen}
  (file open) before it can read or write a file. Note that already at
  this level the association between the word and the action is
  conventional: when one begins working with a file, there is no
  actual \emph{opening} going one simply puts some values in certain
  fields of a structure in central memory, and reads certain data from
  the disk.  From these origins, deep inside the internal working of a
  computer, the term has percolated up to the level of the
  interface.}%
. The expression \dqt{to open a file} in order to change its contents
comes from the programmers' language and its appearence at the level
of the interface--far away from the programming code where it
belongs--represents a serious break of the office metaphor, a break
that is all the worse because the office language already contained
the expression with a different meaning. In a police station, for
example, when one \dqt{opens a file} for a case, one is performing the
action that in a computer is called \emph{create}: the common
expression \emph{open a file} has been, in the computer parlance,
hijacked for a completely different purpose.  The arbitrariety of this
association can be seen further seen in the fact that the icon
represent a folder, which is, \emph{stricti dictu}, not the thing
being opened, while the thing being opened (the document) can't really
be opened in the world of the anchor system.  The arbitrariety of the
association is such that it would be difficult to represent iconically
the computer operation of opening a document. The situation is
depicted in the following diagram:
\begin{center}
\setlength{\unitlength}{0.37pt}
\begin{picture}(528,280)(160,669)
\put(176, 837){\line(1,0){192}}
\put(176, 837){\line(0,1){96}}
\put(176, 933){\line(1,0){192}}
\put(368, 837){\line(0,1){96}}
\put(496, 837){\line(1,0){176}}
\put(496, 837){\line(0,1){96}}
\put(496, 933){\line(1,0){176}}
\put(672, 837){\line(0,1){96}}
\put(368, 885){\vector(1,0){128}}
\put(160, 821){\line(1,0){528}}
\put(160, 821){\line(0,1){128}}
\put(160, 949){\line(1,0){528}}
\put(688, 821){\line(0,1){128}}
\put(496, 669){\line(1,0){176}}
\put(496, 669){\line(0,1){80}}
\put(496, 749){\line(1,0){176}}
\put(672, 669){\line(0,1){80}}
\put(592, 821){\vector(0,-1){72}}
\put(184, 909){picture of}
\put(184, 885){an open}
\put(184, 861){folder}
\put(504, 909){action of}
\put(504, 885){opening a}
\put(504, 861){folder}
\put(376, 893){iconic}
\put(600, 781){symbolic}
\put(504, 725){action of}
\put(504, 701){editing a}
\put(504, 677){document}
\end{picture}
\end{center}
The group formed by the folder icon and the action of opening a
folder is a sign in which the relation between signifier and signified is
at least partially iconic. This whole sign, however, is used as a 
signifier for the action of selecting a document for editing, and the 
relation between this sign-signifier and its signified (what Barthes 
would call the \emph{myth} \cite{barthes:15}) is symbolic.

\bigskip

The third icon is the most unapologetically symbolic of the three
since, in addition to the representational problems pointed out in the
previous case, it breaks the metaphor on which the supposed iconicity
is based. I shall dedicate the next section to the analysis of breaks
in the metaphor, so I shall not consider the issue in depth
here. Suffice it to say that on the actual top of an actual desk there
is no \dqt{saving} any document: one simply stops writing when one is
done. The very use of the term \dqt{saving} tells us more about the
engineering milieu in which these interfaces were designed than about
the action that one accomplishes using this icon, since saving is the
conventional engineering name for the action of storing the contents
of a file on the disk after having copied them in central memory and
modified them.

The aspect that concerns us now is the actual appearance of the
icon. The icon represents a 3 1/2" floppy disk, of a type very common
until about fifteen years ago but that is today virtualyl
disappeared. I can't quite remember the last time I actually saved a
document on a floppy disk rather than on a hard disk or a flash
drive. At one time, the picture had a certain iconic connection with
the action it represented (but not with the anchor metaphor), but that
iconic connection is becoming more and more tenuous, much like the
iconic relation between the Phoenician letter \emph{gimel} and the
back of a camel has been lost in the transition to our letter \dqt{c},
whose relation with its sound is today fully symbolic.

\separate

\begin{figure}
  {\centerline {\immediate\pdfximage height 1cm {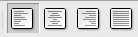}%
      \pdfrefximage\pdflastximage%
    }
  }
  \caption{A series of icons corresponding to functions that to align
    text in a text editing program. The squares reprsent the page, and
    the horizontal lines represent the lines of text. Once the icons
    are understood as a whole, the relation between them and the text
    alignments (left, center, right, justified) is indeed iconic, but
    this iconicity rests on the correct interpretation of the elements
    of the icons, an interpretation that is possible only when the
    icons are presented as a system.}
  \label{align}
\end{figure}

An example of a different nature is given by the four icons of
figure~\ref{align}, which activate the commands to align a portion of
text in a word processor.
The four icons represents a fragment of text aligned on the left,
centered, aligned on the right, and justified, respectively.  The
interesting aspect of these icons is that each one, taken by itself,
is quite obscure (they represent a series of horizontal lines of
varying length) but they become clearer as a group when they are
analyzed differentially. That is, the meaning of each icon depends on
its differential relation with the other icons of the group. No
element, taken in isolation, is a signifier; it becomes one only as a
member of the group, by opposition with the other members, a telling
sign of symbolic reference. True icons signify by themselves, while
symbols signify when they are part of a system of differences without
positive terms \cite{eco:79}. Moreover, the icons are part of a
relatively complex grammar that is, to a large extent, not contained
in the icons themselves.  Consider the possible results of the
selection of the \emph{center} icon in a typical word processor:
\begin{description}
\item[i)] if no text is selected, the paragraph on which the cursor is
  currently placed becomes centered;
\item[ii)] if now we press a \emph{new line} in order to start a new
  paragraph, the new paragraph will also be centerd;
\item[iii)] if a portion of a text is selected, all the paragraphs
  that contain selected text will become centered;
\end{description}
and so on. The \emph{meaning} of the center icon is not determined by
the icon itself, but by its relation with other elements of the
interface (icons, cursors, text,...) at the time of its use. It is, in
other words, defined by a complex grammar of which the icon is but a symbol.

These are but two superficial examples of the type of analysis to
which the elements of a computer interface must be subject in order
to determine their status as elements of communication. In most cases,
the analysis reveals that icons are not icons at all; rather, they
are part of a symbolic system of communication whose effectiveness
depends on a number of (often unspoken) assumptions about the culture
in which the system will be interpreted and used. As such, the iconic
system should be considered tantamount to a language that needs
translation and cultural adaptation no less than any other linguistic
element of the interface.

\section{SOMETHING WEIRD ON THE DESKTOP}
Icons, menus, and labels are the fundamental constituents of the
manifest system, but the metamorphic possibilities of this system rest
on more than just the relation between these elements and the
corresponding elements in the anchor system. For the metaphor to work,
there must be a homomorphic structural relation between the manifest
and the anchor systems. The structural relations between screen
elements must correspond to relation between the corresponding
elements of the anchor system, and actions that alter relations in the
manifest system must correspond to structurally homomorphic actions in
the anchor systems.

In the parlance of interfaces for desktop and laptop computers, the surface of
the screen is called a \emph{desktop}, the files contained in the
computer's disk are \emph{documents}, which can be \emph{opened}
resulting in the appearance of a \emph{window} in which they can be
seen and acted upon. For long documents, the windows only allows the
view of a portion of them, and the visible portion can be changed by
\emph{sliding} the windows up and down the document.  If we take the
metaphor literally, we can observe a few puzzling inconsistencies in
the way it works. One rather evident bizzarry is the screen itself:
desktops are typically horizontal, while the screen is almost always
vertical%
\footnote{At least on desktop computers: tablets are a different
  matter. It is at first slightly puzling that a tablet--which, for
  its horizontal position and mode of interaction would seem the best
  platform to embody a desktop metaphor--has somewhat eschewed the
  desktop. The reason is probably to be found in the origin of tablet
  operating systems: they are by and large adaptations of operating
  systems for telephones, in which the limited size of the screen
  makes the implementation of a desktop metaphor problematic.}%
. Moreover, there is an amount of forcedness in the way every data
file in a computer is metamorphized into a document. This
metamorphosis might have been adequate for early office machines such
as the ALTO, which dealt almost exclusively with written texts and
office drawings, but these days we put many more things in computers
than we did 30 years ago, and many of the things are not immediately
identifiable with documents, at least at the metaphorical level%
\footnote{Of course, everything in a computer is stored in a file, and
files are sequential organizations of data composed of a finite
alphabet and, therefore, can be considered as documents. The computer
file, however, is a programming concept, not an interface one (in the
office world, for instance, a file is a collection of related
documents, a connotation that collides with that of most computer
files); the file-as-a-document can be a useful metaphor for
programmers, but not for the users of the interface, and should stay
hidden in the internal organization of the programs, where it
belongs.}%
.  
In our daily life, we do not ordinarily consider a song or
a video as documents. A document is, in its most common acceptation,
something written on paper (a legal document, for example) or, in
another acceptation---already metaphorical---an artifact conveying
information about something (\dqt{this vase is a document of the
sophisticated pottery technique of the Anasazi indians}). It is
important to realize that, in the latter case, it is not the object
\emph{per se} that is a document, but a particular function that the
object performs in a given situation. Objects \emph{per se} are never
documents. To assimilate the variety of conceptual artifacts that can
be put into a computer with documents stretches the metaphor beyond
its point of usefulness and represents an improper combination of
abstraction and metaphor. I shall return to this problem later in this
section.

The directories of the operating system are modeled as \dqt{folders},
a metaphor with which people accustomed to work in an office are quite
well acquainted, but that will not be as familiar to the majority of
people. The arbitrariety of this choice can be highlighted by
considering an alternative. If files are \dqt{stuff} (documents and
otherwise), then directories are containers of stuff, and the
containers with which most people are acquainted are \emph{boxes}.
This alternative metaphor has a number of advantages: boxes can
contain other (smaller) boxes, while folders do not in general contain
other folders (putting a folder into another folder typically causes a
mess, since all folders are more or less the same size).  We are used
to put a plethora of different things into boxes (while typically we
only put pieces of paper into folders). The boxes metaphor may work
better for some people than the folder metaphor, which is probably
best reserved to computers used in offices, with the caveat that
directories can be nested, while folders typically are a \dqt{flat}
archival mechanism.  New metaphors may require or suggest new
operations. For instance, one can break a box, destroying it and at
the same time \emph{spilling} all its contents into the box that
contained it. In the algebra of the operating system, this operation
corresponds to a local, one-level, flattening of the directory tree.
The point here is not that an anchor system based, say, on a warehouse
with boxes and things inside them is better \emph{per se} than one
based on folders and documets. Quite the opposite, the point is that
there is \emph{no} anchor system that works for all the situations in
which a program might be used. In other words, the anchor system must
be localized whenever it is used in different cultural
communities. Cultural communities do not necessarily coincide with the
national communities to which localization is usually directed: an
American and a French office workers have a lot more in common than an
American office worker and an American fifth-grader, and localization
should take this difference into account%
\footnote{The deleterious consequences of using software prepared for
office workers in educational setting have been amply studied,
especially in connection with presentation software, which may respond
to the needs of corporate presenters, but it certainly doesn't respond
to those of educational institutions \cite{tufte:06}. The point made
here is meant to respond in part, to these concerns.}%
.

In order to be useful, a metaphor must be consistent, at least in its
general lines.  It would be of little use to say that a file is just
like a document if the things one can do in the interface with the
file did not resemble the things that one does with documents. But, of
course, a file is not really a document, so at a certain point the
metaphor will break down. You can't crumple a file in a ball and throw
it in anger in your colleague's face. Not without serious damage to
your computer and to your colleague. You can create a \dqt{shortcut}
to a file, but there is no shortcut to a paper document.  The physical
differences between the two referents of the metaphor, and the failure
of some of the constitutive relations of the anchor system make a
certain degree of metaphor break-up unavoidable.

Nevertheless, some of the metaphor break-ups that one finds in
interfaces could easily have been avoided by a more careful and
informed design. A few examples will suffice. Although word processors
work with documents, the first entry in the menu bar of any word
processor is normally called \dqt{File}, that is, it is named after
the thing that the metaphor is supposed to hide, rather than after the
thing that the metaphor exposes.

Documents of all sorts are \emph{opened} or, in some cases, a
\emph{new} document is created. But in the world outside the computer,
there is no action corresponding to such an abstract notion of
opening.  You \emph{pick up} a written document from a desk, you
\emph{listen} to a song, you \emph{turn on} an apparatus (the obvious
metaphorical base for a program). The idea that the same operation of
opening should have such remarkably different effects (from showing a
written text to playing a song or doing your taxes) is startling. Note
that in order to view a document one does indeed execute a program
but, in this case, the program is outside of the metaphor, and its
presence should be invisible to the user.

The \dqt{new} command is equally perplexing. In an office, one rarely
makes something new. If you want, say to type a letter, you would take
a piece of paper (an existing one), and start writing on it. A more
fitting metaphor in this case would associate to the word processor a
\dqt{paper tray} from which new sheets can be taken to create new
documents.

To make things worse, many of the actions that can be done in modern
programs do not respect the interface metaphor at all: most modern
programs contain a plethora of \dqt{buttons} and \dqt{menus} that, of
course, have no correspondence into the world of the office. The way
we check the spelling of a document in the typical text editing
program, for instance, does not involve taking a dictionary object, as
the metaphor would require, but activating an entry in a rather
counterintuitive menu. In other words, many interface designers have
forgotten that they were supposed to design a direct manipulation
interface to begin with, and transformed it into a forest of buttons
almost as cryptic as an OS/360 JCL script.

\subsection{METAPHOR AND ABSTRACTION}
The previous inconveniences derive from an unfortunate interaction
between metaphor and abstraction. Abstraction (I am simplifying
somewhat) is the practice of ignoring the differences between separate
instances of a situation in order to concentrate on what they have in
common. Abstraction is the main technique to be used in programming
design, but its use in an interface based on metaphor is
problematic. Metaphor is a horizontal relation between two very
concrete and situated sets of elements and actions. Each object in the
anchor system comes to us with its proper type or what Heidegger would
call \emph{readiness-at-hand} \cite{dreyfus:92}, which determines the
specific characteristics of our interaction with that object. A song
is ready-at-hand by being played, a document by being taken, read,
written, torn into pieces, and so on. Abstraction in an interface
destroys our natural interaction with objects and replaces it with
idealized actions that don't fit into the metaphorical relation and
that play against the efforts of the designer to find a metaphor.

There are a number of conclusions that can be drawn from these
considerations. An interface is a semiotic system, and a metaphor is a
relation between two semiotic systems, one of which the intended user of
an application is assumed to be familiar with.  As was already
acknowledged in this paper, if metaphorical interfaces are to work it is
necessary that (1) the anchor system be one with which the intended
user is \emph{really} familiar (not everybody is a silicon valley
geek... not yet, at least) and (2) that the metaphorical relation be
as consistent as possible: every break-up of the metaphor represents a
difficulty for the user of the system.  The consequence of not
respecting these principles are ersatz interaction and obscurity of
usage. 

Years ago, I taught basic computer literacy to adults who had never
used a computer before, and in that occasion I observed quite
consistently that for my students using a modern interface was almost
as hard for them as using a UNIX shell: the metaphor was so unfamiliar
and imperfect as to be virtually of no help. My students simply did
not recognize the office metaphor because none of them had ever worked
in an office, and even when they could recognize some familiar
objects, the behavior of the elements of the interface was so
idiosyncratycally different from what they expected in the
corresponding anchor objects that the metaphor was made completely
useless. Or worse: sometimes using the computer from a command line,
without a metaphor was easier. The initial learning curve was a bit
longer, of course, but once the concepts of file, directory hierarchy,
program, etc. were firmly established, things proceeded more smootly
than with the imperfect graphical interface.

If we use a metaphor for the interface, the metaphor should be
familiar and its use should be consistent. If not, it is probably
better to do without it, which is what some operating systems for very
simple environments are beginning to do.

\subsection{Touch interfaces}
One of the most significant changes in interfaces of the last ten
years has been the introduction of devices with a haptic screen of
small to medium dimensions. The peculiar characteristics of the input device
(e.g. the relative lack of precision in pointing compared to a mouse)
and the small size of the screen (especially in telephones) has pushed
designers towards new solutions, which have generated an interesting
change in the way interfaces are designed.

The most intriguing aspect of this change, from the point of view that
concerns us here, is the great simplification of the anchor
system. The interfaces of modern telephones (and of tablets, which use
the same operating systems) eschew the office metaphor, quite possibly
because a reasonable implementation of it would have been impossible
on the small screen of a telephone, and because the small size and
awkwad typing on the screen keyboard caused the importance of the
written document to wane, and dethroned it from its position as the
centerpiece of the interaction. The metaphor, in this case, is more
playful, almost reminiscent of Aladdin: stroke the right spot on the
screen/lamp, and wonderful things will happen.
\renewcommand{\topfraction}{.99}
\renewcommand{\textfraction}{.01}
\begin{figure}[tbhp]
  \begin{center}
    {\tt
    \setlength{\unitlength}{0.4em}
    \begin{picture}(21,35)(-2,-2)
      \multiput(-1,-1)(20,0){2}{\line(0,1){34}}
      \multiput(-1,-1)(0,34){2}{\line(1,0){20}}
      \put(1,24){
        \multiput(0,0)(0,8){2}{\line(1,0){10}}
        \put(10,0){\line(0,1){8}}
        \put(0,0){\line(0,1){3}}
        \multiput(0,3)(-0.1,0.1){10}{\circle*{0.0000001}}
        \multiput(-1,4)(0.1,0.1){10}{\circle*{0.0000001}}
        \put(5,4){\makebox(0,0){text}}
        \put(0,5){\line(0,1){3}}
      }
      \put(6,19){
        \multiput(0,0)(0,4){2}{\line(1,0){10}}
        \put(0,0){\line(0,1){4}}
        \put(10,0){\line(0,1){1}}
        \multiput(10,1)(0.1,0.1){10}{\circle*{0.0000001}}
        \multiput(11,2)(-0.1,0.1){10}{\circle*{0.0000001}}
        \put(10,3){\line(0,1){1}}
        \put(5,2){\makebox(0,0){text}}
      }
      \put(1,14){
        \multiput(0,0)(0,4){2}{\line(1,0){10}}
        \put(10,0){\line(0,1){4}}
        \put(0,0){\line(0,1){1}}
        \multiput(0,1)(-0.1,0.1){10}{\circle*{0.0000001}}
        \multiput(-1,2)(0.1,0.1){10}{\circle*{0.0000001}}
        \put(0,3){\line(0,1){1}}
        \put(5,2){\makebox(0,0){text}}
      }
      \put(1,9){
        \multiput(0,0)(0,4){2}{\line(1,0){10}}
        \put(10,0){\line(0,1){4}}
        \put(0,0){\line(0,1){1}}
        \multiput(0,1)(-0.1,0.1){10}{\circle*{0.0000001}}
        \multiput(-1,2)(0.1,0.1){10}{\circle*{0.0000001}}
        \put(0,3){\line(0,1){1}}
        \put(5,2){\makebox(0,0){text}}
      }
      \put(6,0){
        \multiput(0,0)(0,8){2}{\line(1,0){10}}
        \put(0,0){\line(0,1){8}}
        \put(10,0){\line(0,1){3}}
        \multiput(10,3)(0.1,0.1){10}{\circle*{0.0000001}}
        \multiput(11,4)(-0.1,0.1){10}{\circle*{0.0000001}}
        \put(10,5){\line(0,1){3}}
        \put(5,4){\makebox(0,0){text}}
      }
    \end{picture}
    }
  \end{center}
  \caption{A schematic view of the interface of the message
    interchange application \emph{Whatsapp}, very popular in modern
    telephones. The text is organized in two columns of \dqt{bubbles}
    of the types used in cartoons to signify the utterances of the
    characters. The icons are divided in two columns, enforcing a
    geometric metaphor of conversation as opposition, related to some
    extent to the metaphor \dqt{debate is war}, at least in the
    geometric analogy with the position of two armies prepared for
    battle.}
  \label{whatinterface}
\end{figure}
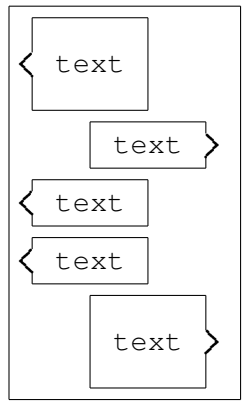

The simplification of the anchor system is obtained thanks to the
possibilities of direct manipulation offered by the haptic screen, as
well as the simplified structure of the opetating system. In these
conditions, the anchor system can be reduced to a minimum: the
interface is basically a bunch of buttons organized in a stack of thin
sheet that can be slided%
\footnote{The sheets are thin because all the things
that we can slide with a finger are thin; the notion of a button on a
sliding sheet is a bit of a stretch, but it seems to create no
problems to most people.}%
. One just pushes a button and launches an application. This
simplification comes at a cost: things work well as long as one use
simple applications that do not interchange data, that is, as long as
the data are hidden inside the applications; manipulating and managing
many documents on a tablet can be a frustrating experience.

While the interface of the operating system has been simplified to the
maximum, the interface of many applications still relies on many
culturally determined models. Consider one of the most popular
applications for modern telephones: the message interchange program
\emph{Whatsapp}. The schema of the interface for a typical
conversation is shown in figure~\ref{whatinterface}, where the blobs
on the left contain incoming messages, and those on the right contain
outgoing ones.
This bilateral organization is not neutral: it reflects a metaphorical
view of a conversation as a spatial arrangements of opposites, the
metaphor that lies behind expressions such as \dqt{take sides} in a
conversation or \dqt{the distance} between two positions.  It would be
interesting to analyze the consequences of organizing an instrument
for friendly chats along such oppositionals lines, but this not the
purpose of this paper. We shall simply note here that the metaphor
along which the conversation is organized is cultural, and it is
reasonable to assume that other cultures, used to different
metaphorical arrangements of conversations (such as a circle of
conversation in which there are no opposites) would be more
comfortable with a different metaphor. Note also that the tips on the
sides of the blocks denote speech in the conventional semiotic system
of cartoon and are therefore highly cultural.

It is at least possible that the optimal adaptation of this program to
different cultures will require translating not only the linguistic
element, but also aspects of the semiotic system. Once again, we are
in the presence of a case in which localization involves more than
just labels and messages, and internationalization should consequently
take a bigger r\^ole in the design of the system.

\section{DESIGN FOR LOCALIZATION}
If we recognize that the interface is a semiotic system akin to a
language (in the communicative sense, not necessarily in the
structural one), then, in order to adapt to various cultural
communities, it must undergo a translation not too different from that
that the labels and messages undergo when we adapt it to different
language communities. In other words, the process of localization
should not be limited to the translation of the interface language,
but it should involve either changing the anchor system to adapt it to
different cultures, or choosing an anchor system that is truly, from
an anthropological point of view, universal. The manifest system will
follow suit, and will be translated as part of the normal localization
process.

All this implies that most of the elements that compose an interface
should not be an unchaengable part of a software system, but should be
a configurable part of it, much like the captions in the menus and the
explanations in the windows.  The \dqt{resident} interface of a
program will, in this model, consist in a series of abstract
operations that determine the general information exchange of the
\emph{conversation} between the system and the user, without
determining the details of the semiotic system in which this exchange
will take place.

Consider, as an example, the problem of finding \emph{350, Willow
lane} in an unknown city. The abstract part of the conversation
include more or less the following:

\begin{description}
\item[i)] I communicate to a person that I need help;
\item[ii)] the person tells me that he is willing to help me;
\item[iii)] I exaplain that I am looking for an address;
\item[iv)] the person might claim that he is not able to help me in
this general problem (for instance, the person also is from out of
town), in which case the conversation ends, or that he can help me;
\item[v)] I communicate the address I am looking for;
\item[vi)] the person might claim that he is not able to help me on
the specific request (for instance, he is from the town, but doesn't
know the address), in which case the conversation will end, or that he
can help me, in which case he will explain how to arrive to the
address I am looking for.
\end{description}

This is an abstract flow of conversation, which can (and in general
is) be implemented using a variety of semiotic systems. I can ask for
help simply by lowering the window of my car and saying \dqt{excuse
me} to somebody (an ambiguous message that is made explicit by the
context); the person may communicate his willingness to help simply by
coming near me. I can receive the indications in English, Franch,
Finnish or whatnot; if I don't speak the local language the person
might draw a route on the map, starting by pointing to a location on
the map and then to me to say \dqt{we are here.} There are endless
possibilities to implement this conversation, depending on the
specific communicative context.

\begin{figure}[hbtp]
  \begin{center}
    {\tt
      \setlength{\unitlength}{5mm}
      \begin{picture}(16,10.5)(0,0)
        \newsavebox{\fsquare}
        \newsavebox{\hsquare}
        \savebox{\fsquare}{
          \multiput(0,0)(0,3){2}{\line(1,0){4}}
          \multiput(0,0)(4,0){2}{\line(0,1){3}}
        }
        \savebox{\hsquare}{
          \put(0,2.5){\line(0,1){0.5}}
          \put(0,3){\line(1,0){4}}
          \put(4,3){\line(0,-1){3}}
          \put(4,0){\line(-1,0){0.5}}
        }
        \put(0.5,0.5){\usebox{\fsquare}}
        \multiput(1,1)(0.5,0.5){3}{\usebox{\hsquare}}
        \put(0.5,6.5){\usebox{\fsquare}}
        \put(7.5,6.5){\usebox{\fsquare}}
        \put(11.5,6.5){\usebox{\fsquare}}
        \put(4.5,9){\vector(1,0){3}}
        \put(7.5,7){\vector(-1,0){3}}
        \put(2.5,3.5){\vector(0,1){3}}
        \put(5,0){\vector(1,1){2}}
        \thicklines
        \multiput(0,6)(0,4){2}{\line(1,0){16}}
        \multiput(0,6)(16,0){2}{\line(0,1){4}}
        \put(2.5,8.5){\makebox(0,0){external}}
        \put(2.5,7.5){\makebox(0,0){interface}}
        \put(9.5,8.7){\makebox(0,0){internal}}
        \put(9.5,8){\makebox(0,0){interface}}
        \put(9.5,7.2){\makebox(0,0){(API)}}
        \put(13.5,8.5){\makebox(0,0){program}}
        \put(13.5,7.5){\makebox(0,0){core}}
        \put(2.5,2.5){\makebox(0,0){messages}}
        \put(2.5,1.5){\makebox(0,0){and labels}}
        \put(11,5){\makebox(0,0)[l]{static portion}}
        \put(11,4.3){\makebox(0,0)[l]{of the program}}
        \put(7,1){\makebox(0,0)[l]{language versions}}
      \end{picture}
    }
  \end{center}
  \caption{Organization of a program from a point of view of standard
    internationalization. The architecture of the program is pretty much
    fixed at the production site and can't be changed. The only changes
    allowed are in the captions file, which is translated in the
    language of the target culture.}
  \label{organization1}
\end{figure}
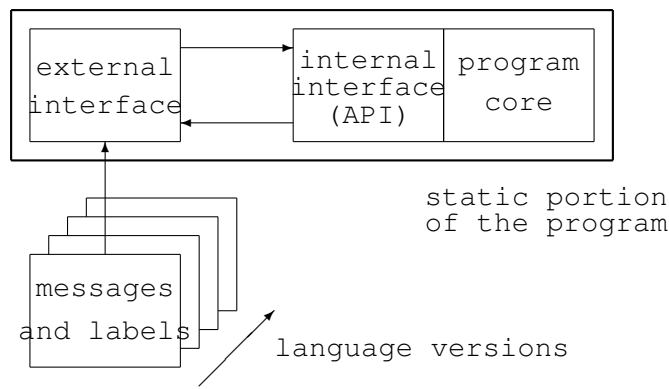

The same should hold for the type of communication represented by a
program's interface. While the basic flow of conversation (determined,
for example, by a conversation-theoretical model \cite{wooffitt:05}) is a characteristic
of the program and should be part of it, the metaphorical
model of the conversation, and the specific semiotic system in which
the conversation will be carried out, should vary with the cultural
environment in which the program is used. In technical terms, this
entails that the semiotic system on which the interface is based (the
anchor system) should be part of the configuration of the program, and
should be fixed only when the program is installed. A simple diagram
may be of help here. The typical organization of a program, from the
point of view of internationalization, is today that of
figure~\ref{organization1}.
The basic functions, performed by the program core, come equipped with
an application program interface (API) that constitutes the internal
algebra of the program. This algebra may have several levels, at
different stages of abstraction, but these levels are in general
neither well separated not well documented, and are more of an
internal structure used by the development group than a formal
requirement of the core. The API that is exported and formally defined
in the requirements is the one between the core and the program
interface. This API is the embodiment of the anchor system, that is,
it is based on the office metaphor; its functions are used directly by
another program module: the external interface, which includes the
icons, documents, buttons, etc. with which the user actually
interacts. This whole part of the program is fixed, and it is repeated
unchanged in any locale in which the program is used. Outside of the
executable portion of the program are the files that contain all the
messages and labels of the interface. These files are edited during
localization, and are locale-specific.

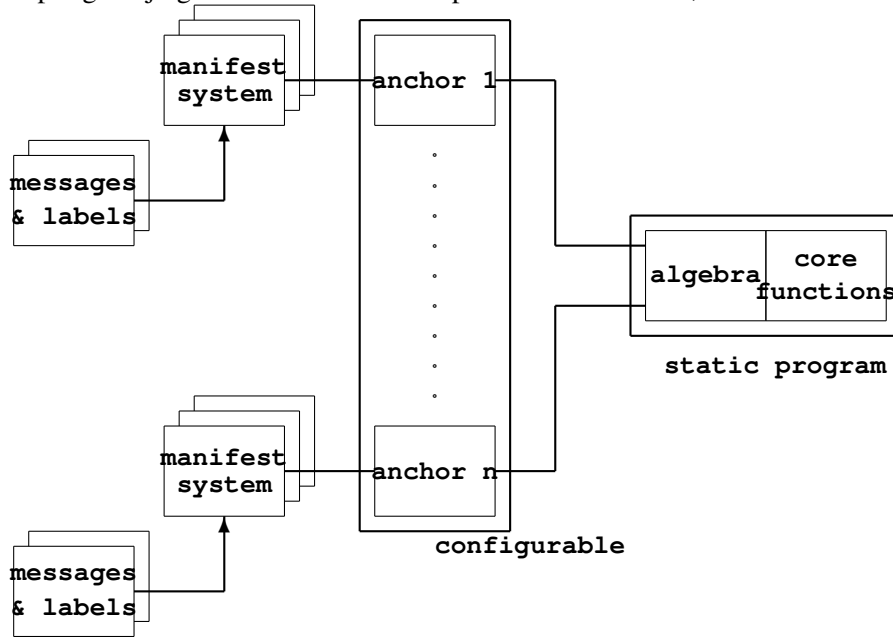
\begin{figure}[tbhp]
  \begin{center}
    {\tt
      \setlength{\unitlength}{4mm}
      \begin{picture}(30,20)(0,0)
        \savebox{\fsquare}{
          \multiput(0,0)(0,3){2}{\line(1,0){4}}
          \multiput(0,0)(4,0){2}{\line(0,1){3}}
        }
        \savebox{\hsquare}{
          \put(0,2.5){\line(0,1){0.5}}
          \put(0,3){\line(1,0){4}}
          \put(4,3){\line(0,-1){3}}
          \put(4,0){\line(-1,0){0.5}}
        }
        \put(0,0){\usebox{\fsquare}}
        \put(0.5,0.5){\usebox{\hsquare}}
        \put(5,4){\usebox{\fsquare}}
        \multiput(5.5,4.5)(0.5,0.5){2}{\usebox{\hsquare}}
        \thicklines
        \put(4,1.5){\line(1,0){3}}
        \put(7,1.5){\vector(0,1){2.5}}
        \thinlines
        \put(0,13){\usebox{\fsquare}}
        \put(0.5,13.5){\usebox{\hsquare}}
        \put(5,17){\usebox{\fsquare}}
        \multiput(5.5,17.5)(0.5,0.5){2}{\usebox{\hsquare}}
        \thicklines
        \put(4,14.5){\line(1,0){3}}
        \put(7,14.5){\vector(0,1){2.5}}
        \thinlines
        \put(12,4){\usebox{\fsquare}}
        \put(12,17){\usebox{\fsquare}}
        \multiput(14,8)(0,1){9}{\circle{0.1}}
        \thicklines
        \put(9,5.5){\line(1,0){3}}
        \put(9,18.5){\line(1,0){3}}
        \thinlines
        \put(21,10.5){\usebox{\fsquare}}
        \put(25,10.5){\usebox{\fsquare}}
        \thicklines
        \put(16,5.5){\line(1,0){2}}
        \put(18,5.5){\line(0,1){5.5}}
        \put(18,11){\line(1,0){3}}
        \put(16,18.5){\line(1,0){2}}
        \put(18,18.5){\line(0,-1){5.5}}
        \put(18,13){\line(1,0){3}}
        \multiput(11.5,3.5)(0,17){2}{\line(1,0){5}}
        \multiput(11.5,3.5)(5,0){2}{\line(0,1){17}}
        \multiput(20.5,10)(0,4){2}{\line(1,0){9}}
        \multiput(20.5,10)(9,0){2}{\line(0,1){4}}
        \put(2,2){\makebox(0,0){\textbf{\small messages}}}
        \put(2,1){\makebox(0,0){\textbf{\small \& labels}}}
        \put(2,15){\makebox(0,0){\textbf{\small messages}}}
        \put(2,14){\makebox(0,0){\textbf{\small \& labels}}}
        \put(7,6){\makebox(0,0){\textbf{\small manifest}}}
        \put(7,5){\makebox(0,0){\textbf{\small system}}}
        \put(7,19){\makebox(0,0){\textbf{\small manifest}}}
        \put(7,18){\makebox(0,0){\textbf{\small system}}}
        \put(14,5.5){\makebox(0,0){\textbf{\small anchor n}}}
        \put(14,18.5){\makebox(0,0){\textbf{\small anchor 1}}}
        \put(14,3){\makebox(0,0)[l]{\textbf{\small configurable}}}
        \put(23,12){\makebox(0,0){\textbf{\small algebra}}}
        \put(27,12.5){\makebox(0,0){\textbf{\small core}}}
        \put(27,11.5){\makebox(0,0){\textbf{\small functions}}}
        \put(29,9){\makebox(0,0)[r]{\textbf{\small static program}}}
      \end{picture}
    }
  \end{center}
  \caption{Organization of a program from a point of view of semiotic
    adaptation as proposed here. The interface rests on an anchor
    metaphor, which can be selected and adapted to the target
    culture. All aspects of the anifest system (text, icons, colors,
    layout, relations with the actions in the anchor system) are
    configured as part of the adaptation of the program to the target
    culture.}
  \label{organization2}
\end{figure}

The analysis carried out in this paper suggests a different structure
(figure~\ref{organization2}).
The static portion of the program, the one that doesn't change with
the locale, includes the core functions and an abstract algebraic
interface. This supports a series of anchor systems, configurable
depending on the locale in which the program is used. The anchor
systems define the metaphorical operations that can be carried
out--basically, they define the whole interface except for the things
that will appear on the screen.  As an example, one anchor system may
allow the creation of a \dqt{new document} the way it is done with
current word processors; another one might place on the screen a ream
of paper from which a new sheet is extracted, and so on. This part of
the system is quite complex from a software development point of view,
and it will require a design effort of considerable sophistication. To
each anchor system correspond a number of possible manifest systems,
that can be selected and configured. Icons, colors, windows (or other
document editing devices), their organization and placement on the
screen should be decided at this level, since they might depend, even
for the same anchor system, on even more stringent requirements of
specific locales (e.g. the position of the title of a window might
depend on the typographic conventions of the target community, and the
position of the OK button may change depending on the writing
direction). 

At this level, we also encounter the standard problems of linguistic
localization: we must adapt the language of the interface to the
different cultural-linguistic communities. This involves linguistic
translation as well as socio-linguistic adaptation: changing the
\dqt{tone} of the language, adjusting the level of formality to the
specific situation (for example: when translating from English to
French, should we use the \emph{tu} or the \emph{vous} form?),
adapting the jargon for localization to specialized audiences, and so on.
\renewcommand{\topfraction}{.99}
\renewcommand{\textfraction}{.01}

\separate

The previous considerations entail a profound structural change, one
that can't simply be reduced to a few guidelines and a few tricks
that make a program more adaptable. The whole process of desiging
complex inter-cultural programs must change. The main change will be that the program
core will no longer include the anchor system and therefore will no
longer accept directly commands from an interface.  Rather, the
program must be thought as a library of functions that communicate
with the interfaces using a \emph{universal metaphor}, implemented as
an algebra of operations, and used by the anchor systems. One
possibility for such universal metaphor is
\emph{conversation}. Conversation is as universal a human activity as
one can find. As universal as language, that is, as universal as
consciousness. Conversation is also quite structured \cite{wooffitt:05}, possibly
structured enough to provide the basis of a formal discipline of
program interfaces. A few mechanisms (turn taking; listing options;
making, accepting, and declining illocutionary statements) preside to
most of it. Conversation analysis is a fairly formal discipline, and
even complex structures such as modality are relatively stable across
cultures \cite{roberts:08}. As a universal foundation for computer interaction,
conversation shows definitely more promise than office work. It is
beyond the scope of this paper to consider the issue of the universal
metaphor any further; the point I should like to make here is that the
native interface of a program should be something natural and highly
abstract. The anchor systems--the metaphors through which the final
user will see the program--should be much more coherent and culturally
dependent.

If we take these considerations to their ultimate consequences,
semiotic localization entails not only a fundamental change in the
structure of a program but also--and most importantly, in the design
process and in the professional figure of the programmer. The current
praxis of internationalization requires a strict (and basically
artificial) separation of two professional cultures. On the one hand,
we have the programmers, technically trained and versed in
mathematics, who design and implement the data structure and the
algorithms necessary to create a compact and efficient core for the
program. On the other hand, we have the translators, teh
anthropologists, the communication theorists or the historians with a
more humanist background, who understand the user communities but lack
the technical savy to intervene on it at a level deeper than that of
message translation.

The program structure proposed here requires a different professional
figure: a humanist with enough computing experience to design and
implement anchor systems. This person might not have a deep knowledge
of mathematics (they won't have to design sophisticated algorithms),
but will have to unite a fairly deep technical knowledge of
programming with a profound knowledge of translation, semiotics,
anthropology, and psychology. It is not too daring to assume that
internationalization and localization should become a college major,
offered jointly by the departments of cultural studies and
computing science.

\section{CONCLUSIONS}
Localization is the adaptation of a program to a homogeneous
(according to some relevant character) class of users; to--the name
says it--a \emph{locale}. But a locale doesn't necessarily correspond
to a linguistic community. A locale may have a cultural identity more
specific than that represented by language or have characteristics
that span a number of linguistic communities. This implies that
localization must be more specific than the simple translation of an
interface's labels and messages: it must encompass the adaptation of all
semiotic systems that make up an interface.  This point of view on localization,
which is essentially a cultural one, has important technical
consequences for internationalization. The direct consequence of our
new point of view is that a program should be customized at a much
deeper level than it was done thus far, and in a more technically
challenging way. While the localization of labels and messages can be
carried out by translators with relatively little technical knowledge,
manipulating files in relatively simple formats using user-friendly
tools, changing the anchor system of a program requires an
intervention of considerable technical sophistication. This, in turn,
requires that a new professional figure, the internationalization
expert, be brought in, with a broad background in languages, semiotics,
anthropology, psychology, and computer science.

When one brings up the kind of considerations that we have done in
this paper to professional meeting, one is met with a whole range of
different reactions. A not uncommone is the \emph{cui prodest} or
even, quite more directly, the \dqt{why bother?}  actitude. In its
most articulated expression, this point of view holds that technology
is a powerful force that unifies cultures so that a culture that uses
computers and wireless telephones is \emph{ipso facto} transformed
into a different culture, one that is either well acquainted with the
way interfaces work or that it better become so. In other words, this
attitude entails that adaptation is a responsibility of the target
culture, not of the designer. The argument, as such, has a few
flaws. The argument is usually not carried out to its final
consequences: if it is the culture that should adapt, why localize at
all? Since most programs are developed in English, leave them in
English and avoid the trouble of different alphabet, Unicode, etc.:
let the user adapt. The argument is seldom taken to these extremes,
and consequently it is left a bit hanging in the air: why should we
adapt certain things but not others? How do we know, from the vantage
point of our culture in which we develop programs, which aspects of an
interface will be important for an unknown target culture?

The issue of who should adapt to what is a complex one, and this is
not the right forum to take it on. We should however at least be open
to the opinion that technology should be at the service of human
culture and not vice-versa. We see a lot of technology that, 
developed in an environment completely alien to that in which it is
deployed, imposes certain habits and mental schemas. The question is
one of humanism and of priorities: is technology a tool or an imposing
culture? Those of us who believe that technology is a tool, that it
should be placed at the service of human values, can't but hope in a
more adaptable and flexible way in which technology can be inserted in
our lives and placed in the service of our values.

These considerations are becoming more pressing with the evolution of
modern technology. If it is true that people who use a computer for
work are more or less immersed in a western-style process of
production and are therefore quite receptive to the office metaphor
behind computer interfaces, it is also true that new devices,
especially telephones, tablets and web pages, are entering in cultural
milieus much more heterogeneous, into ways of life very different from
those of the people who designed these artifacts. Whether technology
will enter these culture by adapting to them, or whether it will
destroy them trying to force them into a predefined mold, is a choice
that computer scientists, as a social group, must make.

\vspace{0.3cm}

\hfill Madrid, May 2018

%\bibliographystyle{newapa} 
%\bibliography{/Users/ssantini/Documents/work/biblio/database} 
%\bibliography{f:/Documents/work/biblio/database} 

\end{document}